\documentclass[10pt]{article}
\usepackage{graphicx, setspace, amsmath, amssymb, wasysym, subfigure, url, multirow, booktabs, verbatim, bm,threeparttable,algorithm, color}
\usepackage{amsthm, rotating}

\usepackage[numbers]{natbib}
\newtheorem{theorem}{Theorem}
\newtheorem{lemma}{Lemma}

\newtheorem{remark}{Remark}

\providecommand{\keywords}[1]{\textbf{\textit{Key words---}} #1}
\makeatletter \def\@biblabel#1{#1.} \makeatother 

\allowdisplaybreaks[3]
\begin{document}
\pdfoutput=1
\setlength{\textheight}{575pt}
\setlength{\baselineskip}{23pt}

\title{Partial Replacement Imputation Estimation Method for Missing Covariates in Additive Partially Linear Model}
\author{Zishu Zhan$^{1,2}$, Xiangjie Li$^3$ and Jingxiao Zhang$^{1,2}$}
\maketitle

\begin{center}

{\it $^1$School of Statistics, Remin University of China\\
	$^2$Center for Applied Statistics, Renmin University of China\\
	$^3$ School of Statistics and Data Science, Nankai University\\
}
\end{center}
\begin{abstract}
	Missing data is a common problem in clinical data collection, which causes difficulty in the statistical analysis of such data. In this article, we consider the problem under a framework of a semiparametric partially linear model when observations are subject to missingness. If the correct model structure of the additive partially linear model is available, we propose to use a new imputation method called Partial Replacement IMputation Estimation (PRIME), which can overcome problems caused by incomplete data in the partially linear model. Also, we use PRIME in conjunction with model averaging (PRIME-MA) to tackle the problem of unknown model structure in the partially linear model. In simulation studies, we use various error distributions, sample sizes, missing data rates, covariate correlations, and noise levels, and PRIME outperforms other methods in almost all cases. With an unknown correct model structure, PRIME-MA has satisfactory performance in terms of prediction, while slightly worse than PRIME. Moreover, we conduct a study of influential factors in Pima Indians Diabetes data, which shows that our method performs better than the other models. 
\end{abstract}


\maketitle

\section{Introduction}
\label{section1}
In many practical situations, the linear model imposes assumptions that are too restrictive to be realistic. Also, the linear model may not be flexible enough to capture the underlying relationship between the response variable and its associate covariates. Furthermore, a linear model can be efficient only if being correctly specified, and otherwise can be seriously biased \cite{Yue2018}. Recently, there has been a rapidly growing literature on the extension of linear regression to tackle the problem of parametric assumptions, including both nonparametric and semiparametric regression. Among them, the additive partially linear model is the most frequently used approach, since it inherits the simple structure of the linear model and retains the flexibility of the nonlinear model. The partially linear model considers only some of the predictors to be modeled linearly, with others being modeled nonparametrically, which allows easier interpretation of the effect of the linear part and may be preferred to a completely nonparametric regression because of the well-known ``overfitting" problem. Since Engle et al. (1986)\cite{Engle1986} introduced the partially linear model to study the effect of weather on electricity demand, the models have been widely studied and extensively used in scientific disciplines, such as epidemiology, medical science, finance, and economics \citep{Speckman1988, Hardle2000, Wang2010}. 

Most studies above are limited to the consideration of complete data cases. In medical research, an investigator's ultimate interest may be in inferring prognostic markers, given the patients' medical predictor variables \cite{Wu2018, Chen2020}. However, data on these covariates are often missing. The most common approach to address missing-data problems is complete-case analysis (CC), which is simple but inefficient. CC can also lead to biased estimates when the data are not missing completely at random. The maximum likelihood method (ML \cite{Dempster1977, Jiang2020}) and the inverse probability weighting method (IPW \cite{Seaman2013, Sun2018}) are also widely used approaches to address missing data. However, likelihood-based methods are sensitive to model assumptions, and re-weighting methods do not always make full use of the available data. Alternatively, imputation \citep{Conn1989, Van2011} is a more flexible approach to couple with missing data. For partially linear models subject to missing data, substantial studies have also been done \citep{Liang2008, Wang2009}. 

Also, due to different data collection procedures, it is common for only a small fraction of records to have complete information and the missing-data patterns are quite complex. For illustration, we provide a data example in Table \ref{tab:example}, in which we just present 10 subjects and 8 covariates and ``$\checked$" means a datum is available. As shown in Table \ref{tab:example}, missing-data patterns vary across subjects, and covariates with missing can exist in both linear and nonlinear parts. Accordingly, new and more capable quantitative methods are needed for complex missing pattern data. Recently, several imputation methods are applicable for individual-specific missing covariates. Lin et al. (2019)\cite{Lin2019} proposed the iterative least-squares estimation (ILSE) method to deal with individual-specific missing data patterns using the classical regression framework, but it may have difficulty accommodating data missing from the partially linear model. Lan et al. (2021)\cite{Lan2021} introduced a model-free semi-supervised learning (SSL) method to impute covariates with high-level missing. However, after imputing missing values, SSL also needs model assumptions to conduct subsequent data analysis such as estimation and prediction. Moreover, most existing methods require the true model structure of the partially linear model. However, in practice, it is difficult to decide which part is really nonlinear and which is not, and incorrect identification of nonparametric and parametric parts may result in substantial biases and poor predictions.

\begin{table}
	\centering
	\caption{A simple example of complex missing patterns in the partially linear model.
		\label{tab:example}}
	\begin{tabular}{c|c|ccc|ccccc}
		\hline
		\multirow{2}*{{\bf Subject}}&\multirow{2}*{${\bm Y}$} & \multicolumn{3}{c|}{{\bf nonlinear part}} &  \multicolumn{5}{c}{{\bf linear part}}\\
		~&~ & $X_1$ & $X_2$ & $X_3$& $X_4$ & $X_5$ & $X_6$& $X_7$ & $X_8$\\
		\hline
		1& \checked &\checked &\checked &\checked &\checked &\checked &\checked &\checked &\checked \\
		2& \checked &\checked &\checked &\checked &\checked &\checked &\checked &\checked &\checked \\
		3& \checked &\checked &\checked & &\checked &\checked & & & \\
		4& \checked &\checked &\checked & &\checked &\checked & & & \\
		5& \checked & \checked &\checked &\checked & &\checked & &\checked &\checked \\
		6& \checked & \checked &\checked &\checked & &\checked & &\checked &\checked \\
		7& \checked &\checked &\checked &\checked & &\checked &\checked &\checked &\checked \\
		8& \checked &\checked &\checked &\checked & &\checked &\checked &\checked &\checked \\
		9& \checked &\checked & & &\checked & &\checked &\checked &\checked\\
		10& \checked &\checked & & &\checked & &\checked &\checked &\checked\\
		\hline
		
	\end{tabular}
\end{table}

In this study, we propose Partial Replacement IMputation Mean Estimation (PRIME), a method that tackles the aforementioned drawbacks of existing methods. The basic idea of PRIME, which is reminiscent of Robinson (1988)\cite{Robinson1988} and Wang (2009)\cite{Wang2009}, is to obtain an estimation by replacing unobserved data with their conditional expectations. Specifically, when the model structure is available, we use spline smoothing to approximate the nonlinear part in the first stage and then replace the unobserved part of the known partially linear model with a kernel estimator mimicking the conditional mean of the unobserved part given the observed part. When the model structure is unavailable, we take a set of semiparametric partially linear models as the candidates, where each sub-model involves only one nonlinear and treats the others as linear. Accordingly, we propose a jackknife weight choice based on complete-case data. Hence, we can combine the sub-models with the weights to obtain averaging final estimators. 
Furthermore, we develop a framework employing projective resampling techniques to address ``the curse of dimensionality" problem in imputation with less information loss. Concretely, we project the covariates along randomly sampled directions to obtain samples of scalar-valued predictors and kernels (dimension reduction). Next, a simple geometric average is taken on the scalar-predictor-based kernel to impute the missing parts (using all-sided information). 

Hence, our method has several advantages, including the following. First, when the model structure is known, PRIME can fully utilize the available information to deal with a high degree of missing data, even data containing no complete observations. Second, the themes we are concerned with here are different from the common methods for partially linear models with complex missing pattern covariates in two aspects: (1) our approach marries the traditionally parallel spline and kernel smoothing techniques, which can tackle the problem of individual-specific missing covariates in the partially linear model, while Wang (2009)\cite{Wang2009} assumes that the missing pattern is not individualized; (2) our method can handle the missing in both linear and nonlinear part, while Wang (2009)\cite{Wang2009} assumes only the linear part can be missing. Third, to the best of our knowledge, our proposed method involving model averaging is the first work to handle the situation with the unknown model structure of the partially linear model.

The remainder of the paper is organized as follows. Section \ref{section2} introduces the basic setup of PRIME with known and unknown model structures, respectively. Theoretical properties are investigated in Section \ref{section3}. Sections \ref{section4} and \ref{section5} present the numerical results using simulated and real data examples, respectively. Section \ref{section6} presents some concluding remarks. Proofs and additional numerical studies can be found in the supporting information. In addition, our proposed method is implemented using R, and the R package {\it misPRIME} to reproduce our results are available at \url{https://CRAN.R-project.org/package=misPRIME}. 

\section{Partial Replacement IMputation Estimation (PRIME)}
\label{section2}
\subsection{PRIME for known model structure}
In this study, we consider the additive partially linear model. let ${\bm Y}=(Y_1,Y_2,\cdots,Y_n)^{\top}$ be the response variable of interest, 
${\bf X}=\{X_{ij}: i=1,2,\cdots,n, j=1,2,\cdots,p+q\}$ be the covariate matrix.
Assume that the random sample $\{(Y_i, {\bm X}_{i}): i=1,2,\cdots,n\}$ is generated by the additive partially linear model:
\begin{eqnarray}\label{gen}
Y_i=\mu_i+\varepsilon_i=\sum_{j=1}^pg_j(X_{ij})+\sum_{j=p+1}^{p+q}X_{ij}{\beta}_j+\varepsilon_i,~~i=1,2,\cdots,n,
\end{eqnarray}
where $\mu_i$ is the conditional mean of $Y_i$ given all the covariates, and $g(\cdot)$ is a smooth function. In our study, we don't require the error term to be Gaussian distributions that much, the $\varepsilon_i$'s can be heteroscedastic. Actually, we only assume that $\varepsilon_i$'s are independently identically distributed random errors, $E(\varepsilon_i|{\bm X}_i)=0$ and $E(\varepsilon_i^2|{\bm X}_i)=\sigma^2({\bm X}_i)$. To prevent ambiguity, the following identification condition are frequently imposed $E[g_j(X_{ij})]=0,~ j=1,\cdots,p$. 

We consider $r_{ij}$ to denote the missing-data indicator for $X_{ij}$, where $r_{ij}$ is 1 if $X_{ij}$ is observable and is 0 otherwise. 
For each unit $i$, $A_i=\{j:r_{ij}=1, \ j=1,2,\cdots,p\}$ and $B_i=\{j:r_{ij}=1, \ j=p+1,p+2,\cdots,p+q\}$ denote the index set of the available covariates, e.g., $A_i=\{1,2,\cdots,p\}$ and $B_i=\{p+1,p+2,\cdots,p+q\}$ represent the complete case. Accordingly, $\bar{A}_i=\{j:r_{ij}=0, \ j=1,2,\cdots,p\}$ and $\bar{B}_i=\{j:r_{ij}=0, \ j=p+1,p+2,\cdots,p+q\}$ denote the index set of the missing covariates. 
Covariates can be divided into four parts based on $A_i$, $B_i$, $\bar{A}_i$ and $\bar{B}_i$: ${\bm X}_{i,A_i}=(X_{ij}: j\in A_i)^{\top}$, ${\bm X}_{i,B_i}=(X_{ij}: j\in B_i)^{\top}$ for observed covariates and ${\bm X}_{i,\bar{A}_i}=(X_{ij}: j\in \bar{A}_i)^{\top}$, ${\bm X}_{i,\bar{B}_i}=(X_{ij}: j \in \bar{B}_i)^{\top}$ for missing covariates. Thus, equation (\ref{gen}) can be expressed as 
\begin{eqnarray*}\label{pred}
	Y_i=\sum_{j\in A_i}g_j(X_{ij})+\sum_{j\in \bar{A}_i}g_j(X_{ij})+\sum_{j\in B_i}X_{ij}{\beta}_j+\sum_{j\in \bar{B}_i}X_{ij}{\beta}_j+\varepsilon_i.
\end{eqnarray*} 
For unobserved part ${\bm X}_{i,\bar{A}_i}$ and ${\bm X}_{i, \bar{B}_i}$, motivated by the core of Robinson (1998) \cite{Robinson1988} and Wang (2009)\cite{Wang2009}, we first define synthetic data $V_{ij}=E[g_j(X_{ij})|{\bm X}_{i,A_i}, {\bm X}_{i, B_i}]$ and $W_{ij}=E[X_{ij}|{\bm X}_{i, A_i}, {\bm X}_{i, B_i}]$, then we have
\begin{equation}\label{cm}
E\left[Y_i-\sum_{j\in A_i}g_j(X_{ij})-\sum_{j\in \bar{A}_i}V_{ij}-\sum_{j\in B_i}X_{ij}{\beta}_j-\sum_{j\in \bar{B}_i}W_{ij}{\beta}_j\Bigg|{\bm X}_{i, A_i},{\bm X}_{i,B_i}\right]=0.
\end{equation}
This implies that  
\begin{eqnarray}\label{cm}
Y_i=\sum_{j\in A_i}g_j(X_{ij})+\sum_{j\in \bar{A}_i}V_{ij}+\sum_{j\in B_i}X_{ij}{\beta}_j+\sum_{j\in\bar{B}_i}W_{ij}{\beta}_j+e_i,
\end{eqnarray}
where $e_i$ are i.i.d. random variables with conditional mean zero given ${\bm X}_{i,A_i}$ and ${\bm X}_{i, B_i}$ for $i=1,2,\cdots,n$.
Hence, by equation (\ref{cm}), we can impute the incomplete part using the information on ${\bm X}_{i,A_i}$ and ${\bm X}_{i,B_i}$. We may write ${\bm U}_{i,C_i}=({\bm X}^{\top}_{i,A_i}, {\bm X}^{\top}_{i, B_i})^{\top}$, $C_i=A_i\cup B_i$ for simplicity. If $g_j(\cdot), j=1,2,\cdots,p$ were known functions, we use the following estimator to ``replace" the missing components of the covariates for unit $i$:
\begin{eqnarray}\label{vhat}
\hat{V}_{ij}=\frac{\sum_{i'=1}^{n}I(C_{i'}\supset \{C_i\cup j\})g_j(X_{i'j}){\bf K}_{\bf H}({\bm U}_{i',C_i}-{\bm U}_{i,C_i})}{\sum_{i'=1}^{n}I(C_{i'}\supset \{C_i\cup j\}){\bf K}_{\bf H}({\bm U}_{i',C_i}-{\bm U}_{i,C_i})}, \quad \left(j \in \bar{A}_i\right),
\end{eqnarray} 

\begin{eqnarray}\label{what}
\hat{W}_{ij}=\frac{\sum_{i'=1}^{n}I(C_{i'}\supset \{C_i\cup j\})X_{i'j}{\bf K}_{\bf H}({\bm U}_{i',C_i}-{\bm U}_{i,C_i})}{\sum_{i'=1}^{n}I(C_{i'}\supset \{C_i\cup j\}){\bf K}_{\bf H}({\bm U}_{i',C_i}-{\bm U}_{i,C_i})}, \quad \left(j\in \bar{B}_i\right).
\end{eqnarray} 
For a given $i$, ${\bf K}_{\bf H}$ is a $m_i$-multivariate kernel function, a probability density that is (typically) symmetric and unimodal at ${\bm 0}$, and that depends on the $m_i\times m_i$ symmetric positive-definite bandwidth matrix ${\bf H}$, where $m_i=|C_i|$ denotes the cardinality of sets $C_i$. A commom notation is ${\bf K}_{\bf H}({\bm U})=|{\bf H}|^{-1}{\bf K}({\bf H}^{-1}{\bm U})$, the so-called scaled kernel, where ${\bf K}(\cdot)$ is also a $m_i$-multivariate kernel function. More details about the choice of kernel function can be found in Section 2.3 and Section 4.

In practice, however, $g_j(\cdot), j=1,2,\cdots,p$ are always unknown. Analogous to Xia et al. (2021)\cite{Xia2021}, we can also approximate the unknown function $g_j(\cdot), j=1,2,\cdots,p$ by polynomial spline-based smoothing. To be specific, we first approximate $g_j(X)$ by a function in polynomial spline space. For simplicity, we assume that all the nonlinear covariates $X_j, j=1,2,\cdots,p$ is distributed on a compact interval $[e_j, f_j], j=1,2,\cdots,p$, and without loss of generality, we take all intervals $[e_j, f_j]=[0,1]$. In practice, for possible $X_j\notin [0,1]$, we can carry out a min-max normalization to $X_j$. Let $U=\{0=$ $\left.{u}_{0}<{u}_{1}<\cdots<{u}_{J_n}<u_{J_n+1}=1\right\}$, where $J_n$ is the number of interior knots increases with sample size $n$. For the $j$-th covariate $X_j$, let ${\bm a}_j(X_{ij})=(a_{j,l}(X_{ij}): l=1,2,\cdots,L)^{\top}$ be the spline basis of degree $d$, where $L=J_n+d+1$. Hence we can write $g_j(X_{ij})$ as ${\bm a}_j(X_{ij})^{\top}{\bm b}_j,$
where ${\bm b}_j=(b_{j,1},b_{j,2},\cdots,b_{j,L})^{\top}$ is the corresponding spline coefficient vector. Accordingly, we can replace the imputed covariate $\hat{V}_{ij}$ by $\hat{\bm a}(X_{ij})^{\top}{\bm b}_j$, where $\hat{\bm a}_j(X_{ij})=(\hat{a}_{j,l}(X_{ij}): l=1,2,\cdots,L)^{\top}$ and 

\begin{equation}\label{ahat}
\hat{a}_{j,l}(X_{ij})=\frac{\sum_{i'=1}^{n} I\left(C_{i'} \supset \{C_i \cup j\}\right) a_{j,l}(X_{i^{\prime} j}) {\bf K}_{\bf H}({\bm U}_{i',C_i}-{\bm U}_{i,C_i})}{\sum_{i'=1}^{n} I\left(C_{i'} \supset \{C_i \cup j\}\right) {\bf K}_{\bf H}({\bm U}_{i',C_i}-{\bm U}_{i,C_i})}, \quad \left(j \in \bar{A}_i\right).
\end{equation}

Applying the above approximation and imputation strategy, we can estimate the unknown parameters in the approximation by the least squares methods. That is, we choose ${\bm b}_1,{\bm b}_2,\cdots,{\bm b}_p$ and $\beta_{p+1},\beta_{p+2},\cdots,\beta_{p+q}$ to minimize the following criterion function
\begin{eqnarray}\label{dls}
\frac{1}{n}\sum_{i=1}^n\left[Y_i-\sum_{j\in A_i}{\bm a}_j(X_{ij})^{\top}{\bm b}_j-\sum_{j\in \bar{A}_i}\hat{\bm a}_{j}({X}_{ij})^{\top}{\bm b}_j-\sum_{j\in B_i}X_{ij}\beta_j-\sum_{j\in \bar{B}_i}\hat{W}_{ij}\beta_j\right]^2,
\end{eqnarray}
and the above least squares problem can be solved directly.

\subsection{PRIME-MA for unknown model structure}
\subsubsection{Candidate model selection}
In practice, the true model structure is always unknown, and it is not easy to identify which part is linear and which is not. This motivates us to use the idea of model averaging to reduce the loss of misspecification of models. With an unknown model structure, we denote the $j$-th candidate model as 
\begin{eqnarray}\label{candt}
Y_i=\mu_{i(j)}+\varepsilon_{i(j)}=g_j(X_{ij})+\sum_{k\neq j}^{p+q}X_{ik}{\beta}_j+\varepsilon_{i(j)},~~i=1,2,\cdots,n.
\end{eqnarray}
Note that each sub-model involves one nonlinear part and $p+q-1$ linear parts. Thus, we can get $p+q$ sub-models. This study shares the same motivation of improving flexibility and prediction accuracy via incorporating one-dimensional marginal regression functions as some existing studies in other contexts \citep{Li2015, Li2018}. For each sub-model, we can obtain $\hat{\mu}_{i(j)}$ using the same imputation strategy mentioned above. Then, we can define the model average prediction of $\mu_i$ as a weighted average of $\hat{\mu}_{i(j)}$. 

\subsubsection{Weight optimization}

An important issue with model averaging is how to choose weights. Here, we propose a weight choice method based on complete-case data. Let the $(p+q)$-dimensional weight vector ${\bm w}=(w_1,w_2,\cdots,w_{p+q})^{\top}$ come from the $\mathcal{R}^{p+q}$ unit hypercube $\mathcal{H}=\left\{{\bm w}\in [0,1]^{p+q}:\sum_{j=1}^{p+q}w_j=1\right\}$. 

Consider the $j$-th model in equation (\ref{candt}) restricted to CC data, we define $S=\{i:C_i=\{1,2,\cdots,p+q\}\}$ as the set of subjects with all covariates observed and $n_0=|S|$ as the sample size of CC data. Hence we can obtain
\begin{equation}\label{cancc}
{\bm Y}_S={\bm \mu}_{S(j)}+{\bm \varepsilon}_{S(j)}={\bf G}_{S(j)}{\bm \gamma}_{(j)}+{\bm \varepsilon}_{S(j)},
\end{equation}
where ${\bm Y}_S=(Y_i:i\in S)^{\top}$, ${\bm \mu}_{S(j)}=(\mu_{i(j)}:i\in S)^{\top}$, ${\bm \varepsilon}_{S(j)}=(\varepsilon_{i(j)}:i\in S)^{\top}$, ${\bf G}_{S(j)}=\left({\bm G}_{1(j)},\cdots,{\bm G}_{n_0(j)}\right)^{\top}$, ${\bm G}_{i(j)}=\left({\bm a}_j(X_{ij})^{\top},{\bm X}_{i(-j)}^{\top}\right)^{\top}$, ${\bm X}_{i(-j)}=(X_{i1},\cdots,X_{i(j-1)},X_{i(j+1)},\cdots,X_{i(p+q)})^{\top}$, ${\bm \gamma}_{(j)}=({\bm b}_j^{\top},{\bm \beta}_{(-j)}^{\top})^{\top}$, ${\bm \beta}_{(-j)}=(\beta_1,\cdots,\beta_{j-1},\beta_{j+1},\cdots,\beta_{p+q})^{\top}$.

For $i \in S$, let $\hat{\mu}^{(-i)}_{(j)}$ be the predicted value of ${\mu}_{i(j)}$ from equation (\ref{cancc}) computed with the $i$-th subject deleted. Define $\hat{\bm \mu}^{cv}_{S(j)}=\left(\hat{\mu}_{(j)}^{(-i)}:i\in S\right)^{\top}$. Then, we can select the optimal ${\bm w}$ by the jackknife version criterion \citep{Hansen2012,Fang2019}
\begin{eqnarray}
CV({\bm w})=\left\|{\bm Y}_S-\sum_{j=1}^{p+q}w_j\hat{\bm \mu}^{cv}_{S(j)}\right\|^2=\left\|\sum_{j=1}^{p+q}w_j\left({\bm Y}_S-\hat{\bm \mu}^{cv}_{S(j)}\right)\right\|^2.
\end{eqnarray}

To tackle the difficulty in caculation of cross-validation estimator, we use the simple relationships ${\bm e}^{cv}_S={\bm Y}_S-\hat{\bm \mu}^{cv}_{S(j)}={\bf D}_{(j)}({\bf I}_{n_0}-{\bf P}_{S(j)}){\bm Y}_S$, where ${\bf D}_{(j)}=diag\left((1-h_{(j)}^{11})^{-1},\cdots,(1-h_{(j)}^{n_0 n_0})^{-1}\right)$, $h_{(j)}^{ii}$ is the $i$-th element of ${\bf P}_{S(j)}$, ${\bf P}_{S(j)}={\bf G}_{S(j)}[{\bf G}_{S(j)}^{\top}{\bf G}_{S(j)}]^{-1}{\bf G}_{S(j)}^{\top}$.
Hence, the cross-validation criterion can be defined as 
\begin{equation*}
CV({\bm w})=\|{\bm e}^{cv}_S{\bm w}\|^2={\bm w}^{\top}{\bm e}_S^{cv\top}{\bm e}_S^{cv}{\bm w}.
\end{equation*}
We then choose the weight vector ${\bm w}$ as follows
\begin{equation*}
\hat{\bm w}^{cv}=\mathop{\arg\min}_{{\bm w}\in\mathcal{H}}{\bm w}^{\top}{\bm e}_S^{cv\top}{\bm e}_S^{cv}{\bm w}.
\end{equation*}
In this way, we can convert the constrained minimization problem into a quadratic programming
problem, which can be implemented using \textit{quadprog} package in R. Denote $\hat{\bm w }^{cv}=(\hat{w}_1,\cdots,\hat{w}_{p+q})^{\top}$. Hence the model averaging prediction of $\mu_i$ can be written as 
\begin{equation*}
\hat{\mu}_i(\hat{\bm w}^{cv})=\sum_{j=1}^{p+q}\hat{w}_j^{cv}\hat{\mu}_{i(j)}.
\end{equation*}

\begin{remark}
	In the case of the unknown model structure, we will focus on the task of using PRIME-MA to do prediction, and the other goals like coefficient estimation will not be much addressed. Actually, in this case, we can not even figure out which part is really nonlinear and which is not, thus coefficient estimation is not feasible.
\end{remark}


\subsection{Computation}
For ease of implementation, one may consider choosing the same kernel function and bandwidth parameter for all equations embedded in equations (\ref{vhat}) and (\ref{what}). In practice, for a given $i$, we assume that ${\bf K}(\cdot)=\prod_{j=1}^{m_i}K(\cdot)$, where $K(\cdot)$ is a symmetric univariate kernel. A possible reconciliation procedure is to consider a diagonal bandwidth matrix ${\bf H}=diag(h_1,\cdots,h_{m_i})$, which yields the kernel density estimation employing product kernels ${\bf K}_{\bf H}({\bm U}_{i',C_i}-{\bm U}_{i,C_i})=\prod_{j}^{m_i}K_{h_j}({U}_{i',C_i,j}-{U}_{i,C_i,j})$, where $K_{h_j}({U}_{i',C_i,j}-{U}_{i,C_i,j})=K(({U}_{i',C_i,j}-{U}_{i,C_i,j})/h_j)/h_j$. ${U}_{i',C_i,j}$ and ${U}_{i,C_i,j}$ are the $j$-th components of ${\bm U}_{i',C_i}$ and ${\bm U}_{i,C_i}$, respectively. It should be noted that all models adopt a common bandwidth and we may use a simple bandwidth selection method for simplicity in simulation study.

However, it may be hard to obtain an accurate estimator by a kernel-smoothing procedure in equations (\ref{what}) and (\ref{ahat}) due to the problem of ``the curse of dimensionality''. Hence, we further transform the estimator in equations (\ref{what}) and (\ref{ahat}) using the projective resampling method. The key idea behind projection resampling/random projection was given in the Johnson-Lindenstrauss lemma \citep{Johnson1984}, which preserves pairwise distances after projecting a set of points to a randomly chosen low-dimensional subspace. There are several previous studies on projection resampling/random projection for dimension reduction, including Shi et al. (2010)\cite{Shi2010} for classification, Maillard and Munos (2012)\cite{Maillard2012} for linear regression, and Le et al. (2013)\cite{Le2013} for kernel approximation. For subject $i', i$ and a given $B<\min_i{m_i}$, we project ${\bm U}_{i',C_i}-{\bm U}_{i,C_i}$ onto random directions $\{{\bm v}_{b,C_i} \in \mathcal{R}^{1\times m_i}, b=1,2,\cdots,B\}$, then obtain $B$ kernel values using the resulting scalars $\{({\bm U}_{i',C_i}-{\bm U}_{i,C_i})^{\top}{\bm v}_{b,C_i}, b=1,2,\cdots,B\}$, and finally integrate the $B$ kernels through the geometric mean. For example, for $j \in \bar{A}_i$, $\hat{a}_{j,l}(X_{ij})$ in equation (\ref{ahat}) can be transformed as follows

\begin{eqnarray*}\label{atilde}
	\breve{a}_{j,l}(X_{ij})=\frac{\sum_{i'=1}^{n} I\left(C_{i'} \supset \{C_i \cup j\}\right) a_{j,l}(X_{i^{\prime} j}) \left\{\prod_{b=1}^{B}K_{h}\left[({\bm U}_{i',C_i}-{\bm U}_{i,C_i})^{\top}{\bm v}_{b,C_i}\right]\right\}^{\frac{1}{B}}}{\sum_{i'=1}^{n} I\left(C_{i'} \supset \{C_i \cup j\}\right) \left\{\prod_{b=1}^{B}K_{h}\left[({\bm U}_{i',C_i}-{\bm U}_{i,C_i})^{\top}{\bm v}_{b,C_i}\right]\right\}^{\frac{1}{B}}}, 
\end{eqnarray*}
where ${\bm v}_{b,C_i}$ is a random vector, with each entry $v_{b,C_i,j}$ chosen independently from a distribution $\mathcal{D}$ that is symmetric about the origin with $E(v_{b,C_i,j}^2)=1$. In practice, we usually generate $v_{b,C_i,j}$ from $N(0,1)$ or $U(-1,1)$, but random vectors from other possible distributions in Achlioptas (2003)\cite{Achlioptas2003} can also be used. The bandwidth here is denoted as $h$ instead of $h_j$, since $({\bm U}_{i',C_i}-{\bm U}_{i,C_i})^{\top}{\bm v}_{b,C_i}$ is one-dimensional.






\section{Asymptotic properties}
\label{section3}
Let ${\bm g}_0=(g_{0j}:j=1,2,\cdots,p)^{\top}$, ${\bm \beta}_0=({\beta}_{0j}: j=p+1,\cdots,p+q)^{\top}$ be the true nonlinear functions and the true parameters. We impose the following assumptions:\\
(A1) $\{Y_i, {\bm X}_i\},~ i=1,\cdots,n$ are i.i.d;\\
(A2) $A_i\perp {\bm X}_i$;\\
(A3) The kernel function $K(\cdot)$ is a symmetric density function with a compact support $[0,1]$ and a bounded derivative;\\
(A4) ${\bm \beta}_0\in \mathcal{B}$, where $\mathcal{B}$ is a bouded set;\\
(A5) For a missing data pattern $C$, let $V_j({\bm u}_C)=E(g_j(X_{j})|{\bm U}_C={\bm u}_C)$, $j=1,\cdots,p$, $W_j({\bm u}_C)=E(X_{j}|{\bm U}_C={\bm u}_C), j=p+1,\cdots,p+q$ be the conditional expectation of nonlinear part and linear part, respectively. Define $f_C({\bm u}_C)$ be the density of ${\bm U}_C$. Assume $V_j({\bm u}_C)$, $W_j({\bm u}_C)$ and $f_U({\bm u}_C)$ have continuous second derivatives with respect to ${\bm u}_C$ on the corresponding support;\\
(A6) Given a positive integer $r$ and $v\in(0,1]$ with $d_1=r+v>1.5$. Let $\mathcal{G}$ be the collection of $g$ on $[0,1]$, whose $r$-th derivative $g^{(r)}$ exists and satisfies the Lipschitz condition of order $v$:
\[
\mathcal{G}=\left\{g(\cdot):\left|g^{(r)}(x_1)-g^{(r)}(x_2)\right|\le M|x_1-x_2|^v\right\}, ~\text{for}~x_1, x_2\in [0,1],
\]
where $M$ is a positive constant. We assume that $g_{0j}\in \mathcal{G}, j=1,2,\cdots,p$;\\
(A7) Define ${\bm Z}^{(1)}_i=(X_{ij}, j=1,\cdots,p)^{\top}$ and ${\bm Z}^{(2)}_i=(X_{ij}: j=p+1,\cdots,p+q)^{\top}$. The random vector ${\bm Z}_i^{(2)}$ satisfies that for any vector ${\bm \omega}\in \mathcal{R}^q$
$$
c_0\|{\bm \omega}\|^2_2\le {\bm \omega}^{\top}E({\bm Z}_i^{(2)}{\bm Z}_i^{(2)\top}|{\bm Z}^{(1)}_i={\bm z}_i^{(1)}){\bm \omega}\le C_0\|{\bm \omega}\|_2^2,
$$
where $c_0$, $C_0$ are positive constants with $c_0<C_0$;\\
(A8) The distribution of ${\bm Z}_i^{(1)}$ is absolutely continuous and its density function is bouned away from zero and infinity on $[0,1]^p$;\\
(A9) The number of interior knots $J_n$ satisfies: $n^{1/2d_1}<<J_n<<n^{1/3}$;\\
(A10) Assume $h\to 0$ as $n\to\infty$, $nh^{p+q-1}/(\ln n J_n^2)\to\infty$.

Assumptions (A1)-(A5) are the same as the conditions in Lin et al. (2019)\cite{Lin2019}. Specifically, Assumption (A1) requires under smoothing to obtain a root-$n$-consistent estimator, which is a commonly used regularity assumption in semiparametric regression. Assumption (A2) is about the missing data mechanism and makes sure that the PRIME estimator is consistent. 
Assumptions (A3)-(A5) are standard in nonparametric regression. Assumption (A3) is achieved when the kernel function is the Gaussian kernel, but it is more general. Assumptions (A6), (A8), (A9) and (A10) are common assumptions on the spline basis used in approximating nonparametric coefficient functions \citep{Stone1985,Liu2011}. Assumption (A7) is widely imposed in the partially linear model literature \citep{Liu2011,Deng2010}.

Let $\mathcal{G}_n$ be the space of splines on $[0,1]$, 
according to the spline theory of De Boor (2001)\cite{De1978}, 
For ${\bm g}_0$, we can find and an additive spline function $\dot{\bm g}=(\dot{g}_j:j=1,2,\cdots,p)^{\top}, \dot{g}_j={\bm a}(X_{ij})^{\top}\dot{\bm b}_j\in\mathcal{G}_n$ such that $||\dot{\bm g}^{\top}{\bm 1}-{\bm g}_0^{\top}{\bm 1}||_{\infty}=O_p(L^{-d_1})$, where $\dot{\bm b}_j=(\dot{b}_{j,1},\dot{b}_{j,2},\cdots,\dot{b}_{j,L})^{\top}$. Define $\dot{\bm \beta}=(\dot{\beta}_{p+1},\cdots,\dot{\beta}_{p+q})^{\top}$, where $\dot{\beta}_{p+1},\cdots,\dot{\beta}_{p+q}$ can be obtained by minimizing 
$$
\frac{1}{n}\sum_{i=1}^n\left[Y_i-\sum_{j\in A_i}{\bm a}_j(X_{ij})^{\top}\dot{\bm b}_j-\sum_{j\in \bar{A}_i}\hat{\bm a}_{j}({X}_{ij})^{\top}\dot{\bm b}_j-\sum_{j\in B_i}X_{ij}\beta_j-\sum_{j\in \bar{B}_i}\hat{W}_{ij}\beta_j\right]^2.
$$

\begin{lemma}
	Under assumptions (A1)-(A9), $\dot{\bm \beta}\to {\bm \beta}_0$ in probability as $n \to\infty$.
\end{lemma}

\begin{lemma}
	Under assumptions (A1)-(A9), $\sqrt{n}(\dot{\bm \beta}-{\bm \beta}_0)\to N(0,{\bf A}^{-1}{\bf \Sigma}({\bf A}^{-1})^{\top})$ in distribution as $n\to \infty$, where the expression of ${\bf \Sigma}$ and ${\bf A}$ are given in the supporting information.
\end{lemma}


\begin{theorem}
	Define $\hat{\bm \gamma}=(\hat{\bm b}^{\top}, \hat{\bm \beta}^{\top})^{\top}$ and $\dot{\bm \gamma}=(\dot{\bm b}^{\top},\dot{\bm \beta})^{\top}$, where $\hat{\bm b}=(\hat{\bm b}_1^{\top},\cdots,\hat{\bm b}_{p}^{\top})^{\top}$ and $\dot{\bm b}=(\dot{\bm b}_1^{\top},\cdots,\dot{\bm b}_{p}^{\top})^{\top}$. Under assumptions (A1)-(A9), $\hat{\bm \gamma}\to\dot{\bm \gamma}$ in probability as $n\to \infty.$
\end{theorem}

Lemma 1 shows that the estimation error for $\dot{\bm \beta}$ is at a negligible order. Lemma 2 investigates the asymptotic distribution of $\dot{\bm \beta}$, which is more complicated due to the uncertainty from the kernel estimator. 
Theorem 1 shows that the estimation error between $\hat{\bm \gamma}$ and $\dot{\bm \gamma}$ is quite small. The proofs can be found in the supporting information. 

\begin{remark}
	The missing assumption in (A2) is different from missing completely at random (MCAR), missing at random (MAR) and missing not at random (MNAR). Similar to Lin et al.\cite{Lin2019}, let $\varepsilon_{i0}=Y_i-\sum_{j=1}^pg_{0j}(X_{ij})-\sum_{j=p+1}^{p+q}X_{ij}{\beta}_{0j}$. As shown in Scenario 1 in simulation studies, if $A_i$ depends in part on $\varepsilon_{i0}$, then the Assumption (A2) is weaker than missing completely at random under homoscedastic errors. Under heteroscedastic errors, the Assumption (A2) is MNAR missingness. As shown in Scenario 2, the proposed method also works well for the cases where the MCAR, MAR and MNAR missing exist at the same time. Generally speaking, our proposed method requires Assumption (A2) to be held for asymptotic properties. However, in practice, our proposed method can also have a good shot when the $A_i\perp {\bm X}_i$ assumption is violated.
\end{remark}


\section{Simulation}
\label{section4}
In this section, we consider several simulated scenarios to highlight the properties of PRIME and PRIME-MA in contrast to some other methods. We experimentally investigate the performance of the following methods:

\begin{description}
	\item[PRIME:] the proposed method with known model structure; 
	\item[PRIME-MA:] the proposed method involving model averaging with unknown model structure; 
	\item[ILSE:] the iterative least-square method in Lin et al. (2019)\cite{Lin2019}; 
	\item[SSL:] the semi-supervised learning method in Lan et al. (2021)\cite{Lan2021};
	\item[ML:] the maximum likelihood method proposed in Jiang et al. (2020)\cite{Jiang2020};
	\item[MI:] the multivariate imputation method by chained equations in Van and Groothuis-Oudshoorn\cite{Van2011};
	\item[CC:] the complete-case analysis method.
\end{description}
Note that ILSE and ML were proposed considering the linear regression and can not be easily applied to the partially linear model; hence, we use them directly without taking the nonlinear part into consideration. We acknowledge that there are other approaches such as those in Wang (2009)\cite{Wang2009} that can be used to address the missing-data problem in the partially linear model. However, Wang (2009)\cite{Wang2009} assumes that the missing pattern is not individualized and missing only happens in the linear part. Hence, the methods like Wang (2009)\cite{Wang2009} are not used as the competing methods in our simulations.

We consider the additive partially liner model:
\begin{equation}
Y_i=\mu_i+\varepsilon_i=\sin(2\pi X_{i1})+\sin(\pi X_{i2})+0.5X_{i3}^3+\sum_{j=4}^{8}X_{ij}\beta_j+\varepsilon_i,~ i=1,\cdots,n,
\end{equation}
where ${\bm \beta}=(1,-1.5,1,-1.2,0.4)^{\top}$. We generate $(X_{i1},X_{i2},X_{i3})$ independently from uniform distribution ${\rm U}[0,1]$ and $(X_{i4},\cdots,X_{i8})$ from multivariate normal distribution $N_{5}({\bf 1},\Sigma)$. For element of $i$-th row and $j$-th column of $\Sigma$, $\rho_{ij}$ represents the correlation between $X_{i,i+3} $ and $X_{i,j+3}$ with three situations: $\rho_{ij}=0.3$, $\rho_{ij}=0.6$ and $\rho_{ij}=0.8^{|i-j|}$. We conduct simulation with $n=200, 400$. Also, similar to Fang et al. (2019)\cite{Fang2019}, we consider the following error distributions: homoscedastic case with $\varepsilon_i\sim N(0,\sigma^2)$; heteroscedastic case with $\varepsilon_i\sim N(0,\sigma_i^2)$, where $\sigma_i^2=\sigma^2 \sum_{j=1}^{p+q}X_{ij}^2/{\rm E}(\sum_{j=1}^{p+q}X_{ij}^2)$. The $\sigma^2$ changes with $R^2=\text{Var}(\mu_i)/\{\text{Var}(\mu_i)+\sigma_i^2\}$. Here, we set $\sigma^2$ so that $R^2=0.7$. 

To evaluate the prediction capabilities of different methods, an independent testing data $\{(\mu_i,{\bm X}_{i}), i=1,\cdots,10000\}$ is added. For methods CC and ML can not predict when missing covariates exist, we generate testing data without missing. The test data is only generated once and remained the same over the simulation runs. For each model setting with a specific choice of parameters, we repeat the simulation $N=1000$ times to evaluate the prediction error (PE) and mean square error (MSE) of the estimator of ${\bm \beta}$ to assess the performances, these are
\[
{\rm PE}=\frac{1}{N}\sum_{l=1}^{N}\frac{1}{n_{test}}\sum_{i=1}^{n_{test}}(\hat{\mu}_i^{(l)}-\mu_i)^2,
\]

\[
{\rm MSE}=\frac{1}{N}\sum_{l=1}^{N}\sum_{j=p+1}^{p+q}(\hat{\beta}_j^{(l)}-\beta_j)^2,
\]
where $\hat{\mu}_i^{(l)}$ is the estimate of $\mu_i$ and $\hat{\beta}_j^{(l)}$ is the estimate of $\beta_j$ in the $l-$th replication. Here, we set $p=3$ and $q=5$ in the simulation study. The MSE based on $N$ replications can be partitioned into MSE=Variance+Bias$^2$, that is
\[
\frac{1}{N}\sum_{l=1}^{N}\sum_{j=p+1}^{p+q}(\hat{\beta}_j^{(l)}-\beta_j)^2=\frac{1}{N}\sum_{l=1}^{N}\sum_{j=p+1}^{p+q}(\hat{\beta}_j^{(l)}-\bar{\hat{\beta}}_j)^2+\sum_{j=p+1}^{p+q}(\bar{\hat{\beta}}_j-\beta_j)^2,
\]
where $\bar{\hat{\beta}}_j=\frac{1}{N}\sum_{l=1}^{N}\hat{\beta}_j^{(l)}$.

For better comparison, we also represent the results of ``PE ratio" for $R^2=0.1,0.3,0.5,0.7,0.9$, which is calculated by dividing the PE of one estimator by the PE of PRIME. A PE ratio that is greater than 1 indicates that the method produces a larger PE than PRIME. 

For the setting of the number of knots, we can use the function {\it{bs($\cdot$,df)}} in R software to generate a B-spline basis matrix, where the number of knots can be specified by setting the degree of freedom argument ``{\it{df=3+number of knots}}". For simplicity, we set {\it{df=3}} to avoid overfitting effect and high-dimensional problems. For the kernel function, based on the Silverman’s rule-of-thumb method, the bandwidth we select is $h_j=1.06\hat{\sigma}_{ U_{C_i,j}}n^{-1/5}$ and the Gaussian kernel is adopted here, where $\hat{\sigma}_{U_{C_i,j}}$ is the standard deviation of the sample of observations of $U_{C_i,j}$. In our simulations, we adopt this common bandwidth for PRIME, PRIME-MA, ILSE and SSL methods.

\subsection[Scenario 1]{Scenario 1: $A_i$ depend on $\varepsilon_{i0}$}

For the missing pattern, refering to Fang et al. (2019)\cite{Fang2019}, we divide the covariates in all cases into 4 groups. Specifically, the first group ${\bm X}_i^{(1)}$ consists of $(X_{i1},X_{i2})$, the second group ${\bm X}_i^{(2)}$ consists of $(X_{i3},X_{i4})$, the third group consists of $(X_{i5},X_{i6})$ and so on. For each sample, variables in the first group ${\bm X}_i^{(1)}$ are always available. In this scenario, we set missing probability of the $i$-th unit as $P({\bm X}_i^{(2)}\text{is missing})=\{1+\exp(a\varepsilon_i+b)\}^{-1}$, $P({\bm X}_i^{(3)}\text{is missing})=\Phi(c\varepsilon_i+d)$ and $P({\bm X}_i^{(4)}\text{is missing})=e$ for the covariates in group 2, 3 and 4, respectively. For missing rate (denoted as MR), there are two settings are considered, where $(a, b, c, d, e)=(0.1, 0.5, 0.1, -1.1, 0.3)$, and $(a, b, c, d, e)=(0.1, 0.3, 0.1, -0.5, 0.6)$, in which the missing rate is about 60\% and 85\%, respectively. 

\begin{figure}[H]
	\begin{center}
		\includegraphics[width=12cm,height=9cm]{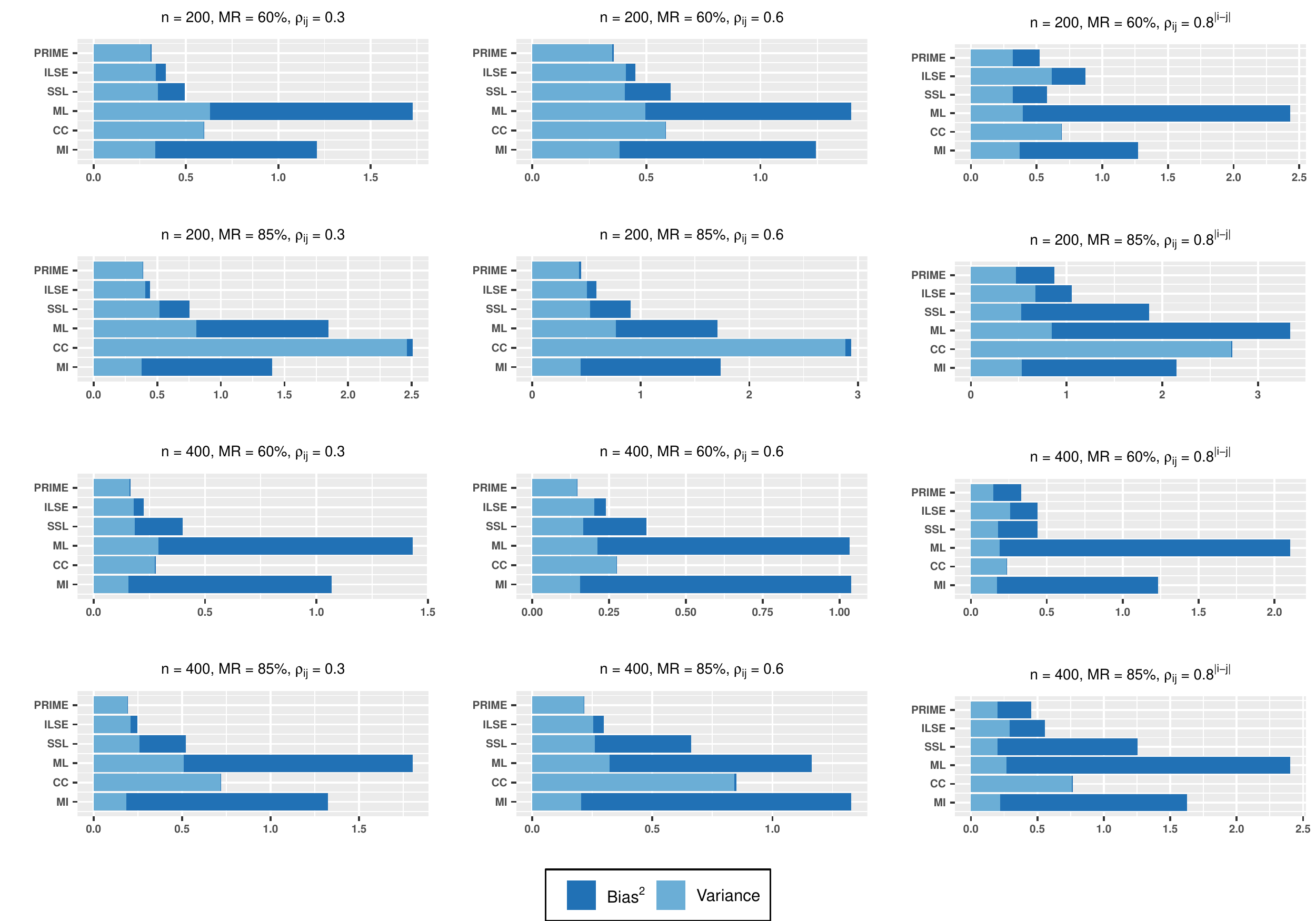}
	\end{center}
	\caption{MSE for different methods in homoscedastic case under Scenario 1.}
	\label{fig:homoMSEm1}
\end{figure}
\begin{figure}[H]
	\begin{center}
		\includegraphics[width=12cm,height=9cm]{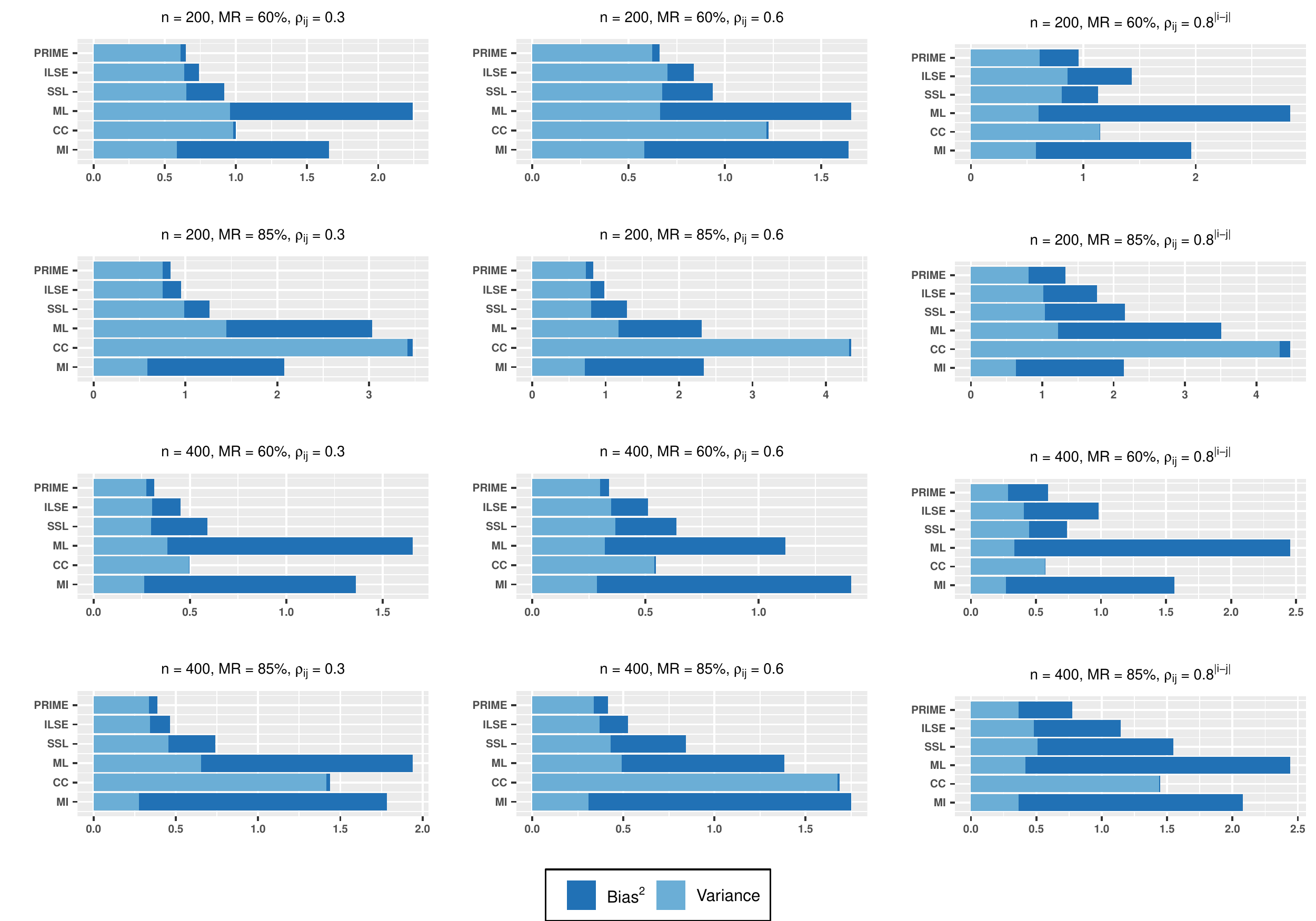}
	\end{center}
	\caption{MSE for different methods in heteroscedastic case under Scenario 1.}
	\label{fig:heteMSEm1}
\end{figure}
\begin{figure}[h]
	\begin{center}
		\includegraphics[width=8cm,height=6cm]{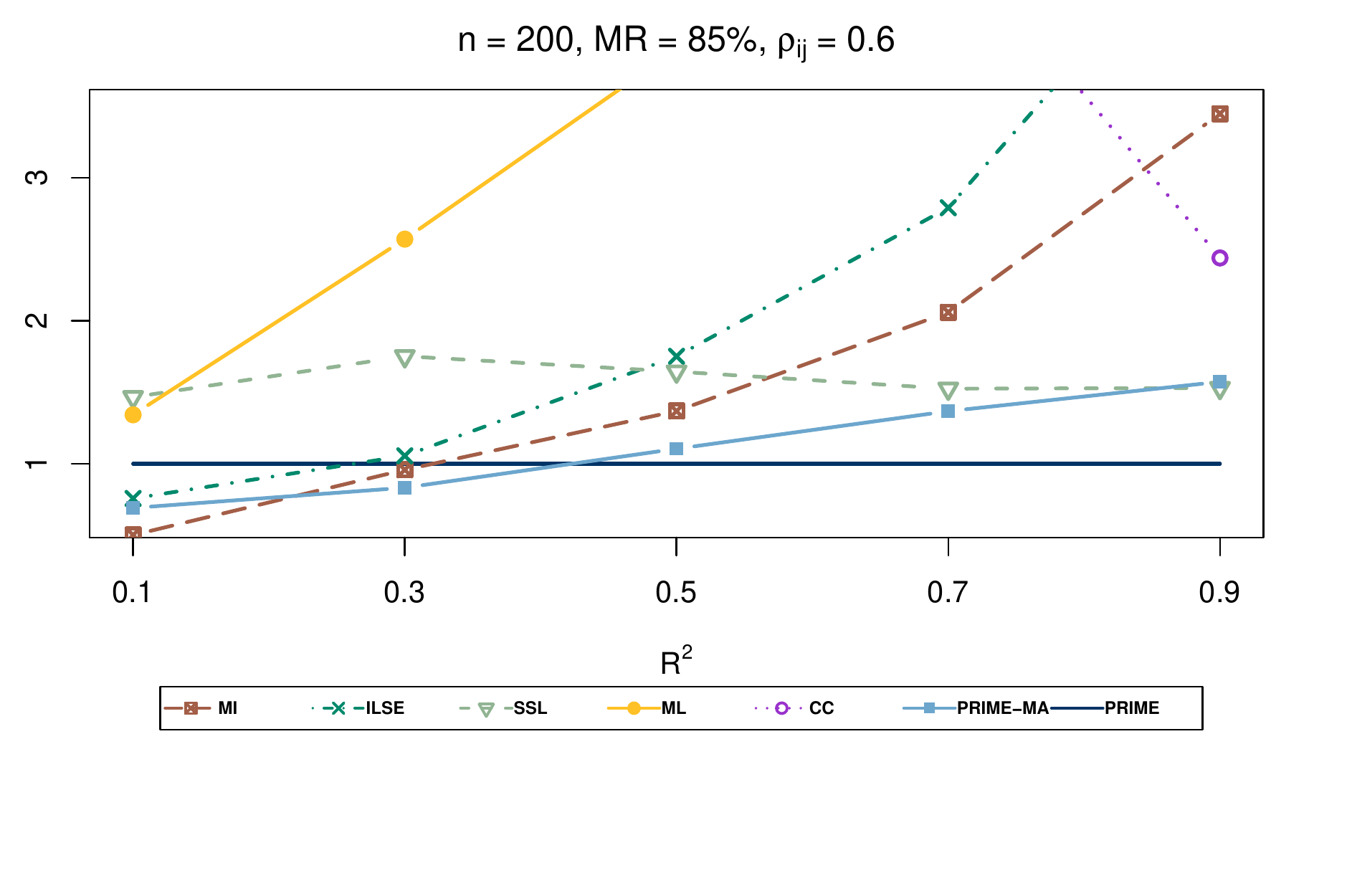}
	\end{center}
	\caption{PE ratio for different methods in homoscedastic case under Scenario 1.}
	\label{fig:homoR2m1}
\end{figure}
\begin{figure}[h]
	\begin{center}
		\includegraphics[width=8cm,height=6cm]{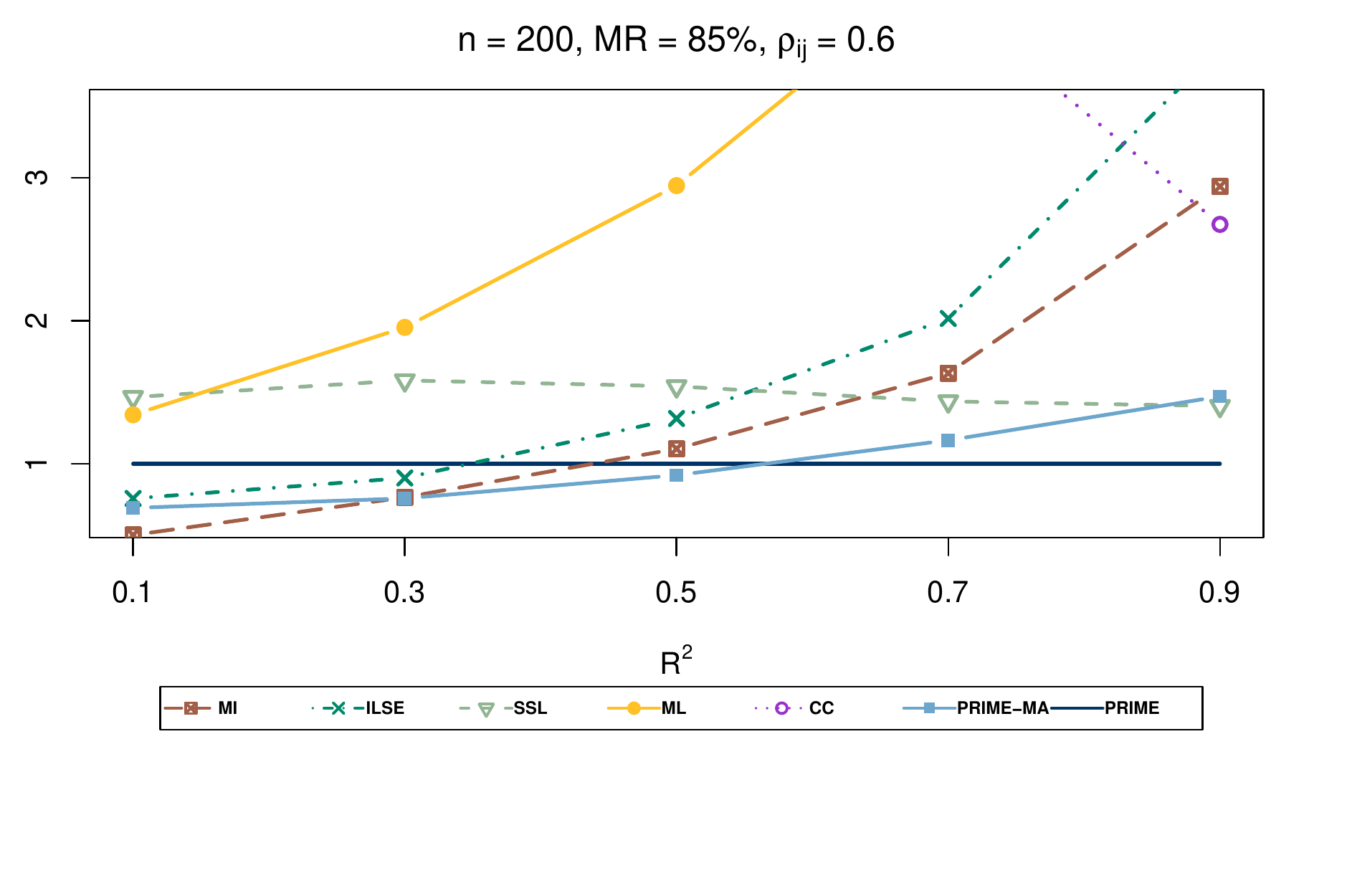}
	\end{center}
	\caption{PE ratio for different methods in heteroscedastic case under Scenario 1.}
	\label{fig:heteR2m1}
\end{figure}

The results of PE are presented in Tables \ref{tab:homoPEm1} and \ref{tab:hetePEm1}. We present the results of MSE in Figures \ref{fig:homoMSEm1} and \ref{fig:heteMSEm1}. Also, Figures \ref{fig:homoR2m1} and \ref{fig:heteR2m1} show the results of ``PE ratio" for $R^2=0.1,0.3,0.5,0.7,0.9$. Due to page restrictions, we only present the results of cases $n=200$, $\rho_{ij}=0.6, \text{MR}=85\%$ for PE ratio. As mentioned above, PRIME-MA is not included in the comparison results of MSE due to the fact that PRIME-MA is used for prediction. The main conclusions are as follows:

\begin{description}
	\item[(i)] Tables \ref{tab:homoPEm1} and \ref{tab:hetePEm1} show that the proposed method PRIME enjoys advantages in prediction accuracy over other methods since it produces the smallest PE in almost all situations. SSL also has satisfactory performance in terms of PE but is worse than PRIME. The PEs of all methods generally increase as expected when the missing rate increases and the sample size decreases, t. Also, the heteroscedastic case yields larger PE than the homoscedastic case.
	
	\item[(ii)] The MSEs of all methods generally increase when the missing rate increases and the sample size decreases, t. The figure shapes with $n=200$ and $n=400$ are different. The comparison results between different methods show that the proposed method PRIME has an advantage over the other five methods when $n=200$. However, when $n=400$ and ${\rm MR}=60\%$, CC approaches to or performs a little better than PRIME. It indicates that CC method may be good enough when dealing with relatively large sample size and low missing rate cases. ML leads to a larger bias than other methods because ML estimators can be highly biased when the missing mechanism assumption and normal distribution assumption do not hold. Not surprisingly, although CC has the smallest estimation error in almost cases, its estimation variance is quite wild and causes trouble in terms of the MSE when $n=200$. 
	
	\item[(iii)] In almost all ranges of $R^2$ considered, PRIME powerfully dominates the other methods except when $R^2$ is small. It makes sense that it is more challenging to describe the nonlinear relationship when the noise level is high. However, in practice, we are not so hung up about the low signal-to-noise ratio case. When $R^2$ increases, the relative performance of different methods has changed. CC’s prediction error is larger when $R^2$ is not very high. The lines of CC and ML do not appear entirely in some figures because these methods lead to much higher PEs than others. Also, the line's tail of CC declines as $R^2$ decreases, indicating the reduced difference between CC and PRIME. ILSE and MI share similar results: the PE ratio is below one in the beginning but increases generally (passing the PRIME line eventually) as the $R^2$ increases. The switching can happen before $R^2=0.5$. The PE ratio of SSL declines slightly when $R^2$ increases, which exhibits that the advantage of PRIME over SSL is less obvious with a high signal-to-noise ratio.
	
	\item[(iv)] As mentioned above, with the known model structure, PRIME generally performs better than other methods. Does our proposed method still work without a true model structure? The answer is yes. PRIME-MA sometimes enjoys second place in terms of PE. It also can be seen from Figures \ref{fig:homoR2m1} and \ref{fig:heteR2m1} that the model averaging method is close or even superior to the PRIME when $R^2$ is less or equal to 0.5, which reconfirms the superiority of the approach of PRIME-MA. In practice, it is difficult to determine the correct model structure when applying the partially nonlinear models. Thus, PRIME-MA is more desirable than other methods because it is relatively easy to implement and has satisfactory prediction accuracy.
\end{description}

\begin{table}[H]
	\centering
	\tiny
	\renewcommand\arraystretch{1.5}
	\caption{PE (SE) with missing data in homoscedastic case under Scenario 1
		\label{tab:homoPEm1}}
	{\tabcolsep=5pt
		\begin{tabular}{@{}cclcccccccc@{}}
			\hline
			{\bf MR} & ${\bm n}$ & ${\bm \rho_{ij}}$ & {\bf PRIME} &{\bf PRIME-MA} & {\bf ILSE} &  {\bf SSL} & {\bf ML}&  {\bf MI}& {\bf CC} \\
			\hline
			\multirow{6}{*}{60\%}  &  \multirow{3}{*}{200} 
			&0.3 & \textbf{0.247} (0.095) & 0.303 (0.116) & 0.657 (0.088) & 0.278 (0.116) & 1.713 (0.952) & 0.490 (0.170) & 0.434 (0.195) \\ 
			&&0.6 & \textbf{0.160} (0.063) & 0.213 (0.082) & 0.643 (0.061) & 0.182 (0.074) & 1.407 (0.549) & 0.316 (0.107) & 0.281 (0.130) \\ 
			&&$0.8^{|i-j|}$  & \textbf{0.099} (0.038) & 0.149 (0.051) & 0.630 (0.044) & 0.105 (0.039) & 2.665 (0.821) & 0.172 (0.056) & 0.141 (0.061) \\ 
			& \multicolumn{1}{l}{\multirow{3}{*}{400}} 
			&0.3 & \textbf{0.125} (0.047) & 0.168 (0.059) & 0.574 (0.042) & 0.170 (0.064) & 1.479 (0.637) & 0.361 (0.102) & 0.202 (0.085) \\ 
			&&0.6 & \textbf{0.078} (0.027) & 0.124 (0.039) & 0.581 (0.031) & 0.108 (0.038) & 1.243 (0.347) & 0.238 (0.069) & 0.127 (0.052) \\ 
			&&$0.8^{|i-j|}$  & \textbf{0.059} (0.019) & 0.103 (0.025) & 0.594 (0.023) & 0.068 (0.022) & 2.594 (0.606) & 0.131 (0.034) & 0.066 (0.025) \\ 
			\hline
			\multirow{6}{*}{85\%}  &  \multirow{3}{*}{200} 
			&0.3 & \textbf{0.374} (0.161) & 0.451 (0.182) & 0.707 (0.112) & 0.495 (0.214) & 2.170 (1.515) & 0.703 (0.206) & 1.771 (1.390) \\ 
			&&0.6 & \textbf{0.247} (0.099) & 0.338 (0.125) & 0.689 (0.085) & 0.377 (0.158) & 1.637 (0.862) & 0.509 (0.151) & 1.166 (1.267) \\ 
			&&$0.8^{|i-j|}$ & \textbf{0.161} (0.063) & 0.255 (0.105) & 0.683 (0.068) & 0.274 (0.104) & 2.983(1.307) & 0.346 (0.092) & 0.537 (0.381) \\ 
			& \multicolumn{1}{l}{\multirow{3}{*}{400}} 
			&0.3 & \textbf{0.184} (0.075) & 0.261 (0.103) & 0.616 (0.059) & 0.310 (0.120) & 1.673 (0.934) & 0.573 (0.145) & 0.564 (0.289) \\ 
			&&0.6 & \textbf{0.118} (0.043) & 0.190 (0.071) & 0.619 (0.044) & 0.261 (0.092) & 1.339 (0.500) & 0.423 (0.093) & 0.341 (0.150) \\ 
			&&$0.8^{|i-j|}$ & \textbf{0.088} (0.029) & 0.149 (0.048) & 0.628 (0.039) & 0.209 (0.063) & 2.685 (0.768) & 0.291 (0.063) & 0.176 (0.081) \\ 
			\hline
	\end{tabular}}
\end{table}

\begin{table}[H]
	\centering
	\tiny
	\renewcommand\arraystretch{1.5}
	\caption{PE (SE) with missing data in heteroscedastic case under Scenario 1.
		\label{tab:hetePEm1}}
	{\tabcolsep=5pt
		\begin{tabular}{@{}cclcccccccc@{}}
			\hline
			{\bf MR} & ${\bm n}$ & ${\bm \rho_{ij}}$ & {\bf PRIME} &{\bf PRIME-MA} & {\bf ILSE} &  {\bf SSL} & {\bf ML}&  {\bf MI}& {\bf CC} \\
			\hline
			\multirow{6}{*}{60\%}  &  \multirow{3}{*}{200} 
			&0.3 & \textbf{0.387} (0.169) & 0.442 (0.192) & 0.763 (0.143) & 0.440 (0.207) & 1.874 (1.246) & 0.631 (0.228) & 0.671 (0.356) \\ 
			&&0.6 & \textbf{0.253} (0.112) & 0.317 (0.140) & 0.711 (0.101) & 0.289 (0.132) & 1.454 (0.626) & 0.411 (0.144) & 0.451 (0.243) \\ 
			&&$0.8^{|i-j|}$ & \textbf{0.182} (0.085) & 0.250 (0.125) & 0.698 (0.083) & 0.202 (0.098) & 2.790 (1.022) & 0.242 (0.091) & 0.279 (0.177) \\ 
			& \multicolumn{1}{l}{\multirow{3}{*}{400}} 
			&0.3 & \textbf{0.196} (0.081) & 0.262 (0.106) & 0.643 (0.077) & 0.260 (0.110) & 1.548 (0.702) & 0.471 (0.143) & 0.310 (0.142) \\ 
			&&0.6 & \textbf{0.130} (0.054) & 0.191 (0.077) & 0.628 (0.060) & 0.171 (0.070) & 1.263 (0.398) & 0.306 (0.092) & 0.215 (0.099) \\ 
			&&$0.8^{|i-j|}$  & \textbf{0.102} (0.042) & 0.159 (0.065) & 0.635 (0.047) & 0.115 (0.048) & 2.573 (0.684) & 0.170 (0.052) & 0.134 (0.073) \\ 
			\hline
			\multirow{6}{*}{85\%}  &  \multirow{3}{*}{200} 
			&0.3 & \textbf{0.581} (0.267) & 0.610 (0.240) & 0.862 (0.201) & 0.758 (0.348) & 2.459 (2.058) & 0.902 (0.298) & 2.356 (2.322) \\ 
			&&0.6 & \textbf{0.384} (0.173) & 0.447 (0.170) & 0.775 (0.128) & 0.552 (0.256) & 1.723 (0.934) & 0.627 (0.204) & 1.634 (1.443) \\ 
			&&$0.8^{|i-j|}$  & \textbf{0.268} (0.124) & 0.363 (0.155) & 0.765 (0.110) & 0.416 (0.182) & 3.195 (1.746) & 0.411 (0.127) & 0.992 (0.954) \\
			& \multicolumn{1}{l}{\multirow{3}{*}{400}} 
			&0.3 & \textbf{0.301} (0.133) & 0.383 (0.153) & 0.704 (0.108) & 0.475 (0.196) & 1.781 (1.026) & 0.697 (0.176) & 0.870 (0.512) \\ 
			&&0.6 & \textbf{0.192} (0.084) & 0.281 (0.117) & 0.674 (0.073) & 0.357 (0.144) & 1.449 (0.603) & 0.495 (0.134) & 0.575 (0.352) \\ 
			&&$0.8^{|i-j|}$ & \textbf{0.151} (0.059) & 0.235 (0.091) & 0.690 (0.063) & 0.300 (0.114) & 2.747 (0.916) & 0.339 (0.083) & 0.343 (0.225) \\ 
			\hline
	\end{tabular}}
\end{table}

\subsection[Scenario 2]{Scenario 2: MCAR, MAR and MNAR}

In this scenario, we consider the situation in which MCAR, MAR and MNAR exist at the same time.
For the missing pattern, we divide the covariates in all cases into 4 groups as mentioned in Scenario 1. We set missing probability of the $i$th unit as $P({\bm X}_i^{(2)} \text{is missing})=\{1+\exp(aX_{i1}+b)\}^{-1}$, $P({\bm X}_i^{(3)} \text{is missing})=\Phi(cX_{i3}+d)$ and $P({\bm X}_i^{(4)} \text{is missing})=e$ for the covariates in group 2, 3 and 4, respectively. For missing rate (denoted as MR), there are two settings are considered, where $(a, b, c, d, e)=(0.1, 0.5, 0.1, -1.1, 0.3)$, and $(a, b, c, d, e)=(0.1, 0.3, 0.1, -0.5, 0.6)$, in which the missing rate is about 60\% and 85\%, respectively.  

\begin{figure}[H]
	\begin{center}
		\includegraphics[width=12cm,height=9cm]{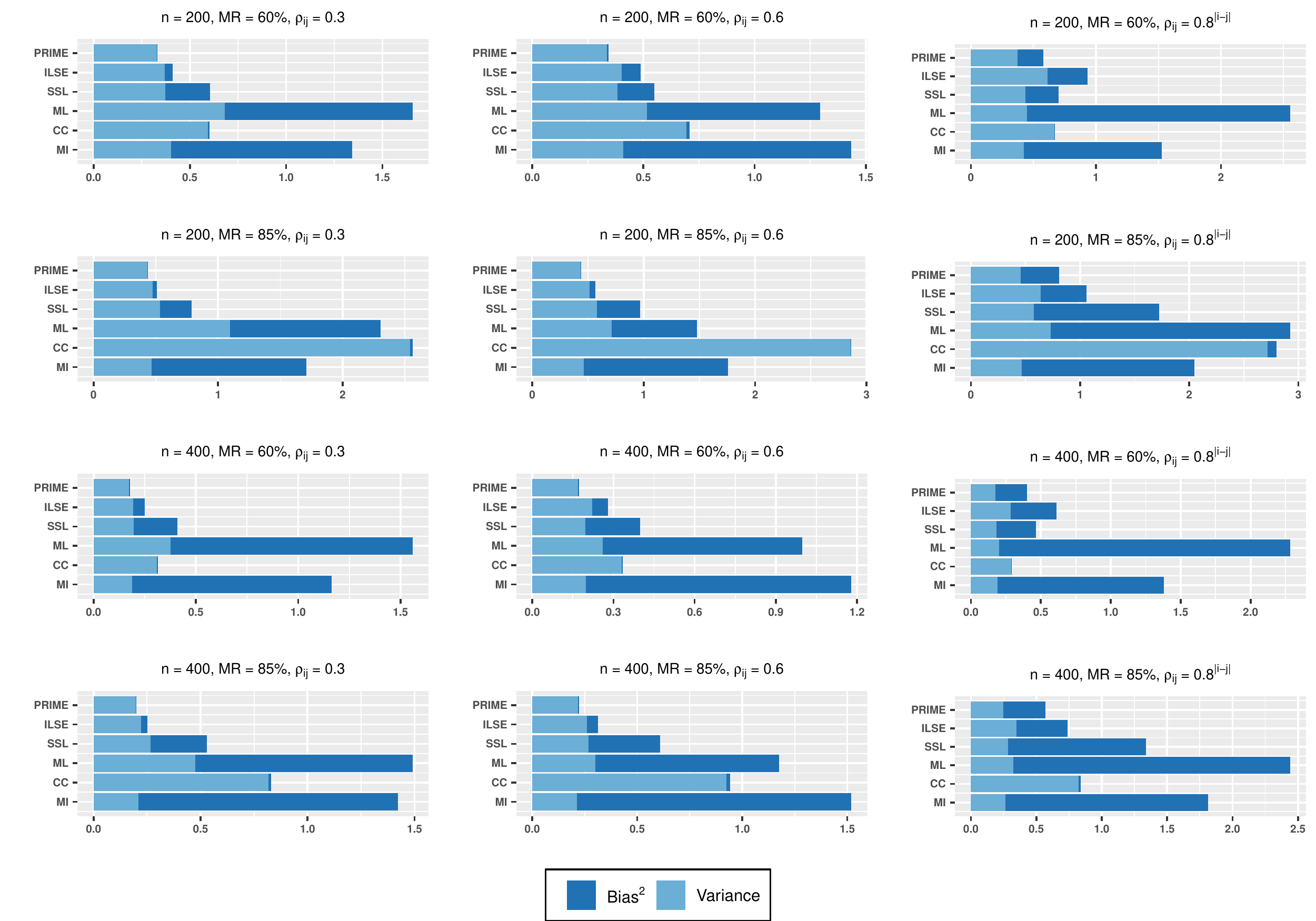}
	\end{center}
	\caption{MSE for different methods in homoscedastic case under Scenario 2.}
	\label{fig:homoMSEm2}
\end{figure}

\begin{figure}[H]
	\begin{center}
		\includegraphics[width=12cm,height=9cm]{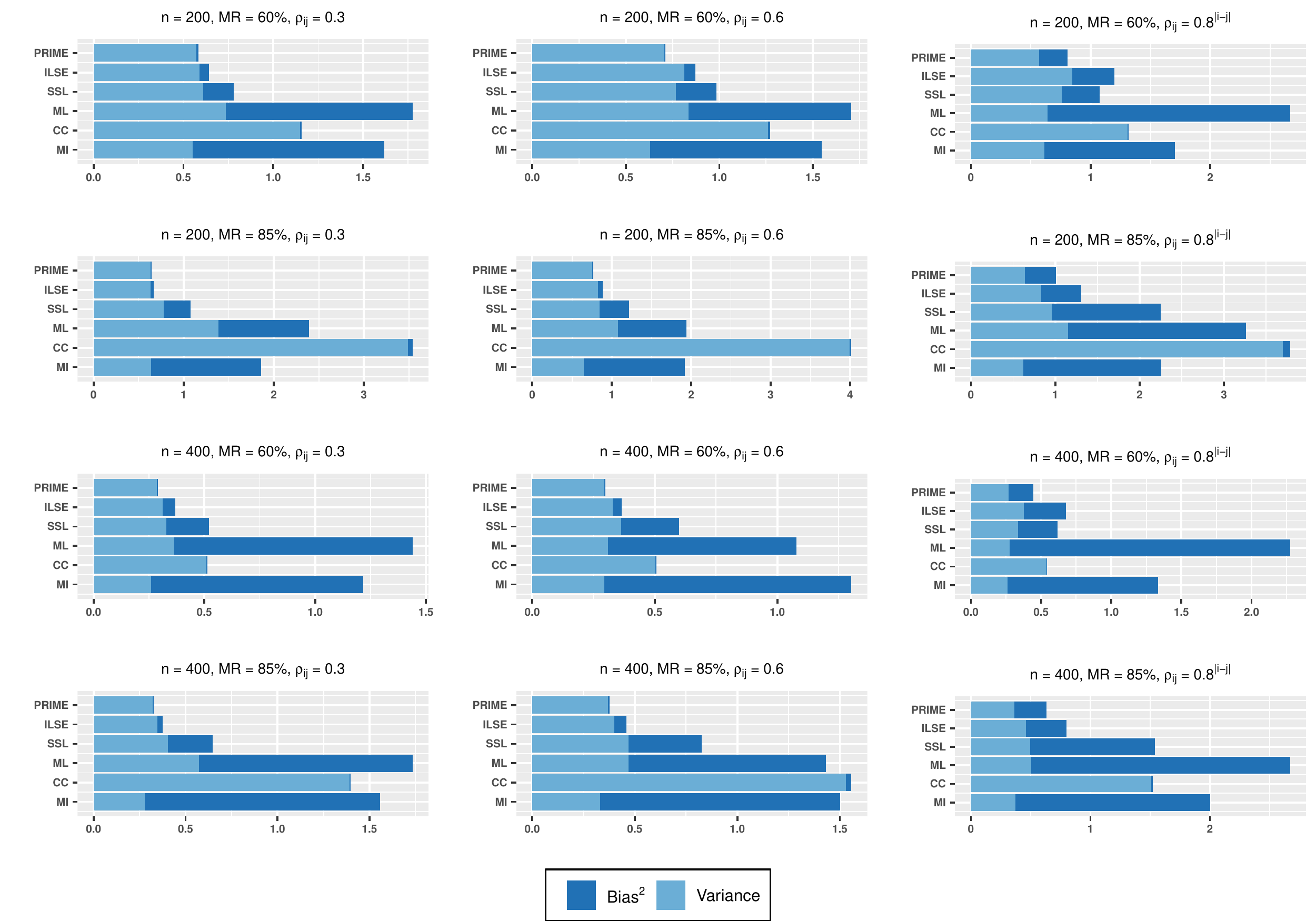}
	\end{center}
	\caption{MSE for different methods in heteroscedastic case under Scenario 2.}
	\label{fig:heteMSEm2}
\end{figure}

\begin{figure}[H]
	\begin{center}
		\includegraphics[width=8cm,height=6cm]{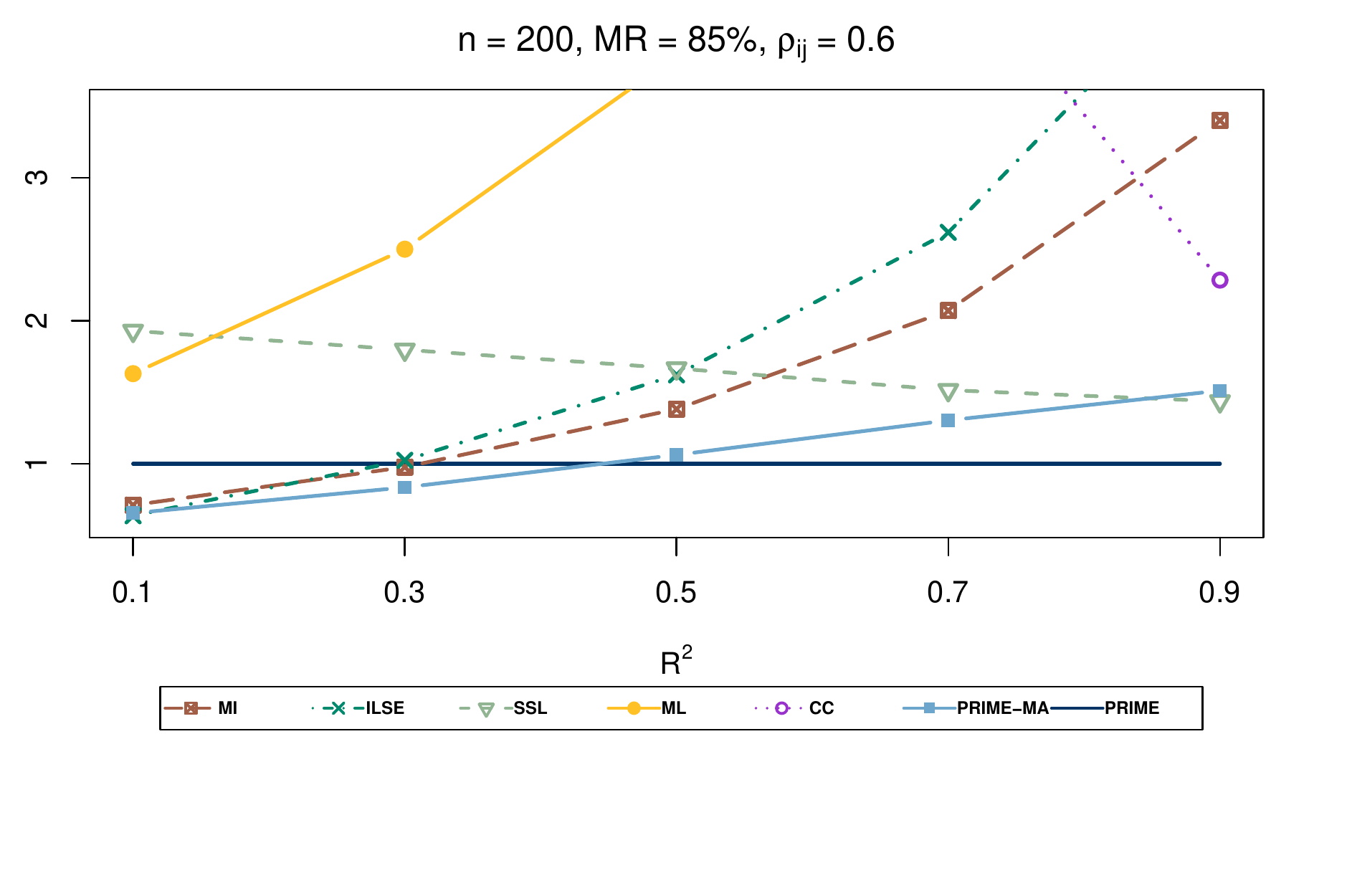}
	\end{center}
	\caption{PE ratio for different methods in homoscedastic case under Scenario 2.}
	\label{fig:homoR2m2}
\end{figure}

\begin{figure}[H]
	\begin{center}
		\includegraphics[width=8cm,height=6cm]{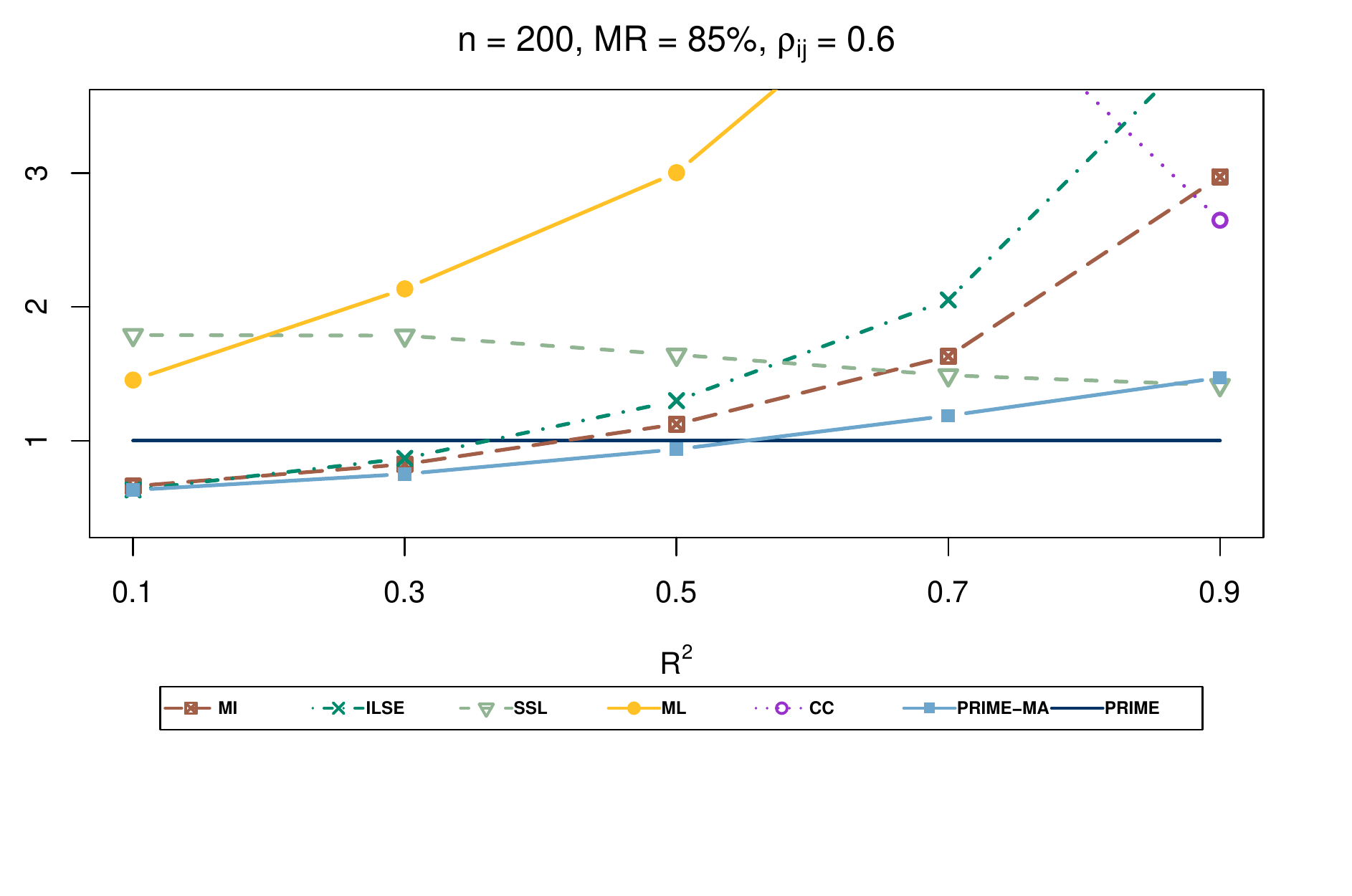}
	\end{center}
	\caption{PE ratio for different methods in heteroscedastic case under Scenario 2.}
	\label{fig:heteR2m2}
\end{figure}

The results of PE are presented in Tables \ref{tab:homoPEm2} and \ref{tab:hetePEm2}. We also display the results of MSE in Figures \ref{fig:homoMSEm2} and \ref{fig:heteMSEm2}. For better comparison, Figures \ref{fig:homoR2m2} and \ref{fig:heteR2m2} represent the results of ``PE ratio" for $R^2=0.1,0.3,0.5,0.7,0.9$. 

Across all results for Scenario 2, PRIME is observed to have superior performance compared to the alternatives in almost all cases, which is analogous to the results in Scenario 1. Specifically, PRIME yields the smallest PE. PRIME-MA and SSL also perform well in terms of PE. ILSE also loses to PRIME-MA, SSL and MI because of mistakenly treating the nonlinear part as the linear part, which indicates that ILSE is not robust against model misspecification. ML, similar to the results in Scenario 1, has the poorest prediction performances. The results in Scenario 2 indicate that PRIME and PRIME-MA can perform well even when the missing mechanism is against the Assumption (A2) mentioned in Section 3.

\begin{table}[H]
	\centering
	\tiny
	\renewcommand\arraystretch{1.5}
	\caption{PE (SE) with missing data in homoscedastic case under Scenario 2
		\label{tab:homoPEm2}}
	{\tabcolsep=5pt
		\begin{tabular}{@{}cclcccccccc@{}}
			\hline
			{\bf MR} & ${\bm n}$ & ${\bm \rho_{ij}}$ & {\bf PRIME} &{\bf PRIME-MA} & {\bf ILSE} &  {\bf SSL} & {\bf ML}&  {\bf MI}& {\bf CC} \\
			\hline
			\multirow{6}{*}{60\%}  &  \multirow{3}{*}{200} 
			&0.3 & \textbf{0.262} (0.102) & 0.308 (0.120) & 0.674 (0.084) & 0.300 (0.115) & 1.699 (1.270) & 0.523 (0.166) & 0.457 (0.204) \\ 
			&&0.6 & \textbf{0.172} (0.063) & 0.223 (0.083) & 0.642 (0.063) & 0.199 (0.073) & 1.392 (1.047) & 0.341 (0.114) & 0.302 (0.118) \\ 
			&&$0.8^{|i-j|}$  & \textbf{0.107} (0.040) & 0.153 (0.053) & 0.629 (0.044) & 0.115 (0.045) & 2.705 (0.895) & 0.187 (0.057) & 0.152 (0.060) \\ 
			& \multicolumn{1}{l}{\multirow{3}{*}{400}} 
			&0.3 & \textbf{0.133} (0.050) & 0.180 (0.062) & 0.600 (0.045) & 0.178 (0.068) & 1.554 (0.702) & 0.399 (0.114) & 0.215 (0.084) \\ 
			&&0.6 & \textbf{0.088} (0.033) & 0.130 (0.041) & 0.575 (0.032) & 0.118 (0.044) & 1.239 (0.360) & 0.258 (0.074) & 0.139 (0.053) \\ 
			&&$0.8^{|i-j|}$  & \textbf{0.063} (0.021) & 0.106 (0.025) & 0.602 (0.026) & 0.074 (0.023) & 2.549 (0.586) & 0.141 (0.039) & 0.071 (0.027) \\ 
			\hline
			\multirow{6}{*}{85\%}  &  \multirow{3}{*}{200} 
			&0.3 & \textbf{0.410} (0.167) & 0.474 (0.181) & 0.728 (0.122) & 0.535 (0.221) & 2.125 (1.575) & 0.773 (0.247) & 1.908 (1.474) \\ 
			&&0.6 & \textbf{0.269} (0.113) & 0.351 (0.128) & 0.705 (0.100) & 0.408 (0.177) & 1.588 (0.862) & 0.558 (0.170) & 1.239 (1.315) \\ 
			&&$0.8^{|i-j|}$ & \textbf{0.172} (0.064) & 0.269 (0.100) & 0.698 (0.069) & 0.287 (0.109) & 3.007 (1.366) & 0.354 (0.099) & 0.625 (0.460) \\  
			& \multicolumn{1}{l}{\multirow{3}{*}{400}} 
			&0.3 & \textbf{0.194} (0.077) & 0.268 (0.105) & 0.606 (0.066) & 0.333 (0.125) & 1.664 (0.957) & 0.623 (0.155) & 0.610 (0.306) \\ 
			&&0.6 & \textbf{0.131} (0.051) & 0.205 (0.082) & 0.629 (0.049) & 0.274 (0.104) & 1.350 (0.524) & 0.441 (0.103) & 0.391 (0.191) \\ 
			&&$0.8^{|i-j|}$ & \textbf{0.093} (0.033) & 0.155 (0.054) & 0.632 (0.041) & 0.221 (0.068) & 2.670(0.799) & 0.303 (0.065) & 0.193 (0.087) \\
			\hline
	\end{tabular}}
\end{table}

\begin{table}[H]
	\centering
	\tiny
	\renewcommand\arraystretch{1.5}
	\caption{PE (SE) with missing data in heteroscedastic case under Scenario 2.
		\label{tab:hetePEm2}}
	{\tabcolsep=5pt
		\begin{tabular}{@{}cclcccccccc@{}}
			\hline
			{\bf MR} & ${\bm n}$ & ${\bm \rho_{ij}}$ & {\bf PRIME} &{\bf PRIME-MA} & {\bf ILSE} &  {\bf SSL} & {\bf ML}&  {\bf MI}& {\bf CC} \\
			\hline
			\multirow{6}{*}{60\%}  &  \multirow{3}{*}{200} 
			&0.3 & \textbf{0.394} (0.102) & 0.447 (0.120) & 0.758 (0.084) & 0.447 (0.115) & 1.819 (1.270) & 0.614 (0.166) & 0.717 (0.204) \\ 
			&&0.6 & \textbf{0.257} (0.063) & 0.310 (0.083) & 0.689 (0.063) & 0.297 (0.073) & 1.472 (1.047) & 0.403 (0.114) & 0.460 (0.118) \\ 
			&&$0.8^{|i-j|}$  & \textbf{0.176} (0.040) & 0.243 (0.053) & 0.679 (0.044) & 0.198 (0.045) & 2.721 (0.895) & 0.236 (0.057) & 0.291 (0.060) \\
			& \multicolumn{1}{l}{\multirow{3}{*}{400}} 
			&0.3 & \textbf{0.194} (0.078) & 0.267 (0.129) & 0.623 (0.066) & 0.253 (0.105) & 1.537 (0.711) & 0.446 (0.138) & 0.337 (0.154) \\ 
			&& 0.6& \textbf{0.124} (0.052) & 0.187 (0.079) & 0.613 (0.048) & 0.167 (0.069) & 1.286 (0.408) & 0.287 (0.088) & 0.212 (0.097) \\ 
			&&$0.8^{|i-j|}$  & \textbf{0.094} (0.038) & 0.156 (0.073) & 0.608 (0.039) & 0.114 (0.046) & 2.593 (0.698) & 0.165 (0.050) & 0.137 (0.070) \\ 
			\hline
			\multirow{6}{*}{85\%}  &  \multirow{3}{*}{200} 
			&0.3 & \textbf{0.549} (0.248) & 0.590 (0.226) & 0.827 (0.165) & 0.739 (0.327) & 2.468 (1.916) & 0.861 (0.293) & 2.536 (2.215) \\ 
			&&0.6 & \textbf{0.369} (0.166) & 0.438 (0.168) & 0.757 (0.124) & 0.550 (0.257) & 1.739 (0.963) & 0.602 (0.190) & 1.695 (1.566) \\ 
			&&$0.8^{|i-j|}$ & \textbf{0.258} (0.117) & 0.360 (0.153) & 0.736 (0.101) & 0.437 (0.196) & 3.178 (1.643) & 0.404 (0.118) & 1.045 (0.952) \\ 
			& \multicolumn{1}{l}{\multirow{3}{*}{400}} 
			&0.3 & \textbf{0.270} (0.113) & 0.369 (0.151) & 0.678 (0.089) & 0.470 (0.184) & 1.794 (1.084) & 0.674 (0.193) & 0.904 (0.523) \\ 
			&&0.6 & \textbf{0.181} (0.078) & 0.269 (0.112) & 0.657 (0.064) & 0.363 (0.147) & 1.392 (0.553) & 0.476 (0.117) & 0.571 (0.311) \\ 
			&&$0.8^{|i-j|}$ & \textbf{0.135} (0.053) & 0.228 (0.095) & 0.661 (0.056) & 0.314 (0.119) & 2.695 (0.978) & 0.336 (0.082) & 0.357 (0.246) \\
			\hline
	\end{tabular}}
\end{table}

\section{Pima Indians Diabetes Database}
\label{section5}
This dataset is originally from the National Institute of Diabetes and Digestive and Kidney Diseases\cite{Smith1988}, which is available at \url{https://data.world/data-society/pima-indians-diabetes-database}. The objective is to predict based on diagnostic measurements whether a patient has diabetes.
Several constraints were placed on the selection of these instances from a larger database. In particular, all patients here are females at least 21 years old of Pima Indian heritage with medical variables: {\it Pregnancies}: Number of times pregnant; {\it Glucose}: Plasma glucose concentration a 2 hours in an oral glucose tolerance test; {\it BloodPressure}: Diastolic blood pressure (mm Hg); {\it SkinThickness}: Triceps skinfold thickness (mm); {\it Insulin}: 2-Hour serum insulin; {\it BMI}: Body mass index; {\it Age}: Age (years) and {\it DiabetesPedigreeFunction}: Diabetes pedigree function. We use the continuous-variable {\it DiabetesPedigreeFunction} as the response. For simplicity, we conduct a standardized transformation to scale response and a min-max normalization to predictors. Furthermore, we exclude subjects with missing {\it DiabetesPedigreeFunction} because the aforementioned methods (PRIME) apply primarily to the missing covariates. Finally, 615 patients are enrolled for statistical analysis. 

To compare the performance of the proposed method with existing methods, we randomly split
the data into a testing set and a training set 100 times. Specifically, each testing set consists of 400 samples randomly selected from the complete observations. The remaining 215 samples constitute the training set, corresponding to an 80\% missing rate of the training set. 

Here, like in the simulation studies, we also set bandwidth based on Silverman's rule-of-thumb. For unknown model structure, we consider the {\it Pregnancies, Insulin} and {\it BMI} to be the nonlinear covariates. We use PE to evaluate the estimation accuracy. However, because of the unknown $\mu$, we are unable to evaluate PE as described in the simulation. Hence, we calculate $\widetilde{\rm PE}$ instead, as follows:

\[
\widetilde{\rm PE}=\frac{1}{N}\sum_{l=1}^{N}\frac{1}{n_{test}}\sum_{i=1}^{n_{test}}(\hat{\mu}_i^{(l)}-Y_i)^2,
\]
where $Y_i$ is the response in testing set and $\hat{\mu}_i^{(l)}$ is the estimate of $\mu_i$.

The PE results are shown in Table \ref{tab:real}. The results show that PRIME-MA has advantages over other methods in prediction as it produces the smallest PE. The missing mechanism and the wrong model structure assumption may give rise to the worse performance of PRIME, ILSE, SSL, ML, MI and CC.

\begin{table}[H]
	\centering
	\tiny
	\renewcommand\arraystretch{1.5}
	\caption{PE (SE) with missing data in real data analysis
		\label{tab:real}}
	{\tabcolsep=5pt
		\begin{tabular}{@{}cclcccccc@{}}
			\hline
			{\bf PRIME-MA} & {\bf PRIME} & {\bf ILSE} &  {\bf SSL} & {\bf ML} & {\bf MI} & {\bf CC}\\
			\hline
			\textbf{1.050} (0.119) &1.226 (0.532) & 0.132 (0.123) & 1.136 (0.121) & 1.420 (0.173) & 1.221 (0.146) & 4.722 (7.701) \\ 
			\hline
	\end{tabular}}
\end{table}




\section{Conclusions}
\label{section6}
In this study,  we propose a method called Partial Replacement IMputation Mean Estimation (PRIME) method to estimate the partially linear regression with complex missing covariates. The core idea of PRIME is to obtain an estimation by replacing unobserved data with their expectations conditional on the observed data. When the model structure is known, we adopt PRIME to do prediction and estimation by marrying the spline and kernel smoothing techniques. When the model structure is unavailable, we take a set of semiparametric partially linear models as the candidates, where each sub-model involves only one nonlinear and treats the others as linear. Then, we apply jackknife model averaging to do prediction. PRIME can fully utilize the available information to deal with a high degree of missing data. Also, our proposed method involving model averaging (PRIME-MA) can handle the situation with the unknown model structure of the partially linear model. Another important advantage our proposed methods offer over existing methods focusing on the partially linear model is their ability to handle missing both in linear part and nonlinear part simultaneously. 

However, the aforementioned work has been developed for a classical setting. Developing PRIME/PRIME-MA to combine a generalized linear model with missing data warrants future research. The distinct natures of the generalized model and linear model present a non-trivial work to obtain a formula like equation (\ref{cm}). Furthermore, we considered only missing covariates even though it is common to encounter cases where both covariates and responses are missing. Hence, developing methods to address practical issues will be the focus of our future work. For the asymptotic properties, when the model structure is known, we focus on the estimation performance on the linear part, while the discussion of the nonlinear part is also important, and deserves further research. Also, in practice, the missing mechanism assumption is too restrictive to be realistic. A less stringent assumption is welcome to provide in the future.
\bibliographystyle{SageH} 

\end{document}